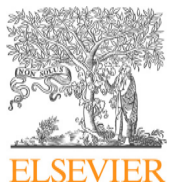
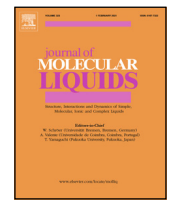

# Charge delocalization and hyperpolarizability in ionic liquids

Carlos Damián Rodríguez-Fernández [a], Luis M. Varela [a], Christian Schröder [b], Elena López Lago [a,*]

[a] *NaFoMat group, Departamento de Física Aplicada and Departamento de Física de Partículas, Universidade de Santiago de Compostela, E-15782 Santiago de Compostela, Spain*
[b] *University of Vienna, Department of Computational Biological Chemistry, Währingerstr. 17, A-1090 Vienna, Austria*

ABSTRACT

In this work the role that charge delocalization plays in the non-linear optical response of ionic liquids is evaluated. The first hyperpolarizability for the non-linear process of second harmonic generation (SHG) and second hyperpolarizability for the non-linear process of electro-optical Kerr-Effect (EOKE) of a large number of ionic liquid forming ions were estimated by means of density functional theory calculations. The results point to that both charge delocalization and molecular geometry are the key features that govern their hyperpolarizabilities. Our findings show that some of the most commonly used anions in ionic liquids are expected to present strong non-linear responses while common cations present a much more limited performance. However, this limitation can be overcome by a proper tailoring of cations to present charge delocalization over large molecular regions. The hypothesis of additivity of hyperpolarizabilities in ionic liquids is tested and exploited to obtain a map of second and third order non-linear susceptibilities of 1496 ion combinations. This map is intended to be a guide for future works on the hyperpolarizability of ILs.

© 2021 The Authors. Published by Elsevier B.V. This is an open access article under the CC BY license (http://creativecommons.org/licenses/by/4.0/).

## 1. Introduction

In recent years optical properties of Ionic Liquids (ILs) have attracted an increasing attention in the field of photonics due to their promising properties and high degree of tunability. ILs are formed by the combination of two different ions that can be individually tailored to control the optical response of the complete IL. In this regard, there are ILs designed to show luminescence in the visible range by introduction of luminescent cations [1], anions [2] or by doping them with transition metal salts [3,4], transition metal complexes [5,6], photoluminescent organic molecules [7] or both [8]. Moreover, by carefully choosing the ions, it is possible to induce the formation of mesophases at a certain temperature range [9,10] or to provide them with thermochromic character [11,12]. Other important optical property that strongly depends on the ion election is refractive index [13], which is the most widely studied optical-related magnitude in ILs, both from experimental [13–16] and computational points of view [17–21].

As other organic molecules, ILs can show different non-linear (NL) optical phenomena, however, the number of works devoted to experimentally determine NL optical magnitudes in ILs is still scarce [22–35]. In contrast, several characterization techniques based on NL optics were successfully applied to ILs for studying their ultra-fast dynamics [36–39], bulk [23,40], or interface [41–48] structure. Recently, some computational works have shed more light in the magnitude of the NL optical response of ILs by calculating the first and second hyperpolarizabilities of some ILs as well as of their forming ions [23,49–51,35,52,53]. However, despite these pioneering works, further investigation on the origins of this NL response is required to determine reliable rules to design ILs with tailored NL optical properties. In this regard, it is well-known that charge delocalization induced by $\pi$-orbitals largely increases the NL response of organic molecules [54,55]. Recent works, such as those of Refs. [56,57], deal with the influence of structure and/or charge delocalization in the NL response of families of organic molecules different from those yielding ILs.

The purpose of this work is to determine the role that delocalized electronic charge in $\pi$-orbitals plays in the NL optical response of a variety of IL-forming ions. For this reason, the molecular volume, first and second hyperpolarizabilities of 92 ions were systematically calculated at the B3LYP/6–311++G(d,p) level of theory for two NL processes, Second Harmonic Generation, SHG, $\beta(-2\omega;\omega,\omega)$, and Electro-Optical Kerr-Effect, EOKE, $\gamma(-\omega;\omega,0,0)$. The results are discussed in terms of the hyperpolarizability and molecular volume ratio of each species, and the most interesting ions for enhancing the non linear response of ILs are identified. Additionally, the hypothesis of additive hyperpolarizability in ILs is revisited and exploited to provide a complete map of the EOKE susceptibilities in the IL world, comprised by 1496 ion combinations. This map is intended to be a valuable tool to guide the future research on the non linear optical properties of ILs.

## 2. Theoretical and Computational Details

### 2.1. Considered Ions

The considered ions include a set of anions that are commonly found in literature, together with cations specially chosen to study the influence of the presence of $\pi$-orbitals in their NL response. The name, abbreviation and structure of all the ions can be consulted in the Supplementary Material. The cations can be classified into three different groups.

The first one comprises nine different families of cations, including ammonium and phosphonium based-cations and seven different cationic heterocycles, all of them bearing an alkyl chain of variable length. While the ammonium-based cations, $[N_{0,0,0,k}]^+$ and $[N_{1,1,1,k}]^+$, and phosphonium-based cations, $[P_{4,4,4,k}]^+$, $[P_{8,8,8,8}]^+$ and $[P_{6,6,6,14}]^+$, present a purely saturated character, the heterocyclic cations present different extents of charge delocalization, ranging from those clearly aromatic such as 1-alkylpyridinium, $[C_k py]^+$, or 1-alkyl-3-methylthiazolium, $[C_k mthia]^+$, to those clearly aliphatic, such as 1-alkyl-1-methylpiperidinium, $[C_k mpip]^+$, or 1-alkyl-1-methylpyrrolidinium, $[C_k mpyrr]^+$.

The second set of cations are singly and doubly substituted imidazolium derivatives incorporating $\pi$-bonds in different extents and positions. Some examples of the attached substituents are allyl, benzyl, crotyl or vinyl groups.

The third set of cations is composed by four different families of imidazoliums that differ in the functionalization and extension of their side chains. It includes a family bearing a perfluorinated alkyl chain, 1-methyl-3-perfluoroalkylimidazolium, $[F_k mim]^+$, a family with a side chain incorporating oxygen, 1-meth(oxyethyl)$_k$-3-methylimidazolium, $[m(eo)_k mim]^+$ and a polyenyl side-chain, $[uC_k mim]^+$, which provides direct access to a charge delocalization region of variable extension. Furthermore, the standardized imidazolium cation bearing an alkyl chain of variable length is also considered, 1-alkyl-3-methylimidazolium, $[C_k mim]^+$.

### 2.2. Hyperpolarizabilities

Polarizability, $\alpha$, SHG first hyperpolarizability, $\beta(-2\omega,\omega,\omega)$, and EOKE second hyperpolarizability, $\gamma(-\omega,\omega,0,0,)$, at $\lambda = 1100$ nm of the considered ionic species and ionic pairs were calculated using DFT at the B3LYP/6–311++G(d,p) level of theory by means of the Gaussian 16 rev. C.01 program [58]. The Coupled-Perturbed Hartree–Fock (CPHF) method [59] was used to obtain the frequency-dependent hyperpolarizability values. All the calculations were performed on properly optimized geometries and, except indicated otherwise, on isolated ions in gas phase. The level of theory was chosen in accordance with that used in recent works dealing with the NL optical properties of ILs, which rely on the same level of theory [35] or combinations of the B3LYP potential with related basis sets [50,49].

As hyperpolarizabilities are tensors whose elements depend on the molecular direction, all the values provided in this work correspond to those more widely accessible from the experimental point of view when dealing with liquid samples. That is, in the case that all the involved fields are parallel polarized, the first hyperpolarizability in the direction ($i$) of the molecular dipole, $\beta_{\parallel}$, and the isotropic average of the second hyperpolarizability, $\langle \gamma \rangle$, [54,60–62]:

$$\beta_{\parallel} = 1/5 \sum_j (\beta_{ijj} + \beta_{jij} + \beta_{jji}) \quad (1)$$

and

$$\langle \gamma \rangle = 1/15 \sum_{i,j} (\gamma_{iijj} + \gamma_{ijij} + \gamma_{ijji}). \quad (2)$$

Finally, molecular volumes of the ions, $V$, were calculated at the same level of theory using the Monte-Carlo procedure implemented in Gaussian, where the molecular volume is defined as the volume inside a contour of an electron density of 0.001 electrons/Bohr$^3$ [63]. Since the output of the algorithm is slightly different after each iteration [64], the presented volumes are those obtained from an average of 100 executions. Polarizabilities, hyperpolarizabilities (in S.I and esu units) and molecular volumes of all the ions studied in this work can be consulted in the Supplementary Material.

Hyperpolarizabilities are magnitudes that increase with the size of a molecule. Since ionic species of very different sizes and features are compared in this work, we decided to discuss them in terms of their hyperpolarizability and molecular volume ratio, magnitude we will refer as molecular hyperpolarizability density. Since both first and second hyperpolarizabilities are considered, we defined both Molecular First Hyperpolarizability Density, M1HD$=\frac{\beta_{\parallel}}{V}$, and Molecular Second Hyperpolarizability Density, M2HD$=\frac{\langle \gamma \rangle}{V}$.

### 2.3. The effect of Charge Delocalization

The influence of charge delocalization in the hyperpolarizability of molecules can be depicted using different quantum–mechanical models [65–68]. In one of the simplest approximations, the delocalized $\pi$ electrons are described as an electron gas of 2N electrons inside a potential well of length $L$ under the perturbation of an external electric field. The second hyperpolarizability of this system can be calculated by summing up the fourth derivative of the energy $E_\xi$ of each excited state $\xi$ in the system with respect to the electric field $F$ [65]:

$$\gamma = -2\sum_{\xi=1}^{N} \frac{1}{6} \frac{\partial^4 E_\xi}{\partial F^4} = \frac{128 L^{10}}{a_0^3 e^2} \sum_{\xi=1}^{N} \left( \frac{-2}{9\pi^6 \xi^6} + \frac{140}{3\pi^8 \xi^8} - \frac{440}{\pi^{10} \xi^{10}} \right), \quad (3)$$

where $e$ is the electron charge and $a_0$ the Bohr radius. According to Eq. (3), the second hyperpolarizability presents a tenth-power dependence on the size of the region where the electrons are delocalized, $L$. This distance can be just the length of a $\sigma$ or a $\pi$ bond, or that of conjugated systems of coupled $\pi$ orbitals. As for second hyperpolarizability, a power dependence on $L$ is also expected for first hyperpolarizability [67,68]. However, additionally, first hyperpolarizability also presents a strong dependence on the geometry of the considered system, since it must be zero in the presence of inversion symmetry [54,69]. Furthermore, some recent studies also reported that, for a specific one dimensional potential well, it is possible to predict a maximum theoretical value for first hyperpolarizability [70] as well as both a minimum and a maximum value of the second hyperpolarizability [71].

In order to evaluate the extent of charge delocalization of the ions considered in this work, we used as a reference the absolute value of the charge-delocalization estimator known as Multi-Center Bond Order (MCBO) index [72]. This parameter was calculated by means of the Multiwfn 3.8 program [73]. The MCBO of each ion was calculated in the two possible directions of the regions where charge delocalization was evaluated. Furthermore, as in our case these regions include different numbers of atoms, we worked with the normalized version of the MCBO parameter [74].

## 2.4. Calculation of non-linear susceptibilities

From a microscopic perspective, an external electric field over a molecule, $F$, induces a non-linear dipole, $\mu_i$, in the direction $i$:

$$\mu_i = \mu_i^{(0)} + \alpha_{ij} \cdot F_j + \frac{1}{2}\beta_{ijk} \cdot F_j F_k + \frac{1}{6}\gamma_{ijkl} \cdot F_j F_k F_l + \ldots, \quad (4)$$

where $\mu_i^{(0)}$ is a possible permanent dipole, and $\alpha_{ij}$, $\beta_{ijk}$ and $\gamma_{ijkl}$ are, respectively, the linear polarizability, first order hyperpolarizability and second order hyperpolarizability. In addition, $i, j, k,$ and $l$ stand for the components on each dimension of the dipole, polarizability and hyperpolarizability tensors. However, from a macroscopic perspective, the NL response of a material is governed by the corresponding nth-order NL susceptibilities [54,75–77], $\chi^{(2)}$ and $\chi^{(3)}$ (in SI units):

$$\chi^{(2)}_{ijk}(-\omega_a;\omega_b,\omega_c) = f(\omega_a)f(\omega_b)f(\omega_c) \cdot \frac{\beta_{ijk}}{\varepsilon_0 V}, \quad (5)$$

$$\chi^{(3)}_{ijkl}(-\omega_a;\omega_b,\omega_c,\omega_d) = f(\omega_a)f(\omega_b)f(\omega_c)f(\omega_d) \cdot \frac{\gamma_{ijkl}}{\varepsilon_0 V}, \quad (6)$$

where $\varepsilon_0$ is the vacuum electric permittivity, $V$ is the molecular volume of the involved molecules and $f(\omega_x)$ is the microscopic field factor at the field frequency $\omega_x$ with $x = a, b, c, d$, the frequencies involved in the NL process. In the Lorentz–Lorenz approximation, this field factor takes the form $f(\omega_x) = (n^2(\omega_x) + 2)/3$, being $n(\omega_x)$ the refractive index of the material at the frequency $\omega_x$, and the product of microscopic field factors can be approximated by an average field factor $f(\bar{\omega})$ at an average frequency $\bar{\omega}$ [76,77]. Approximating the field factors, and using the effective hyperpolarizabilities defined in Eqs. 1 and 2, we obtain:

$$\chi^{(2)}_{IL}(-\omega_a;\omega_b,\omega_c) \approx f(\bar{\omega})^3 \frac{\beta^{IL}_{\parallel}}{\varepsilon_0 V_{IL}}, \quad (7)$$

and

$$\chi^{(3)}_{IL}(-\omega_a;\omega_b,\omega_c,\omega_d) \approx f(\bar{\omega})^4 \frac{\langle\gamma^{IL}\rangle}{\varepsilon_0 V_{IL}}, \quad (8)$$

where $V_{IL}$, $\beta^{IL}_{\parallel}$ and $\langle\gamma^{IL}\rangle$ are the molecular volume and hyperpolarizabilities of the IL.

**Table 1**
Normalized MCBO indices, M1HD and M2HD of ammonium, phosphonium and heterocycle-based cations bearing the same alkyl chain length ($k = 4$). For the [C$_4$quin]$^+$ cation the MCBO was calculated on the perimeter involving both cycles.

| Cation | |MCBO| | M1HD | M2HD |
|---|---|---|---|
|  |  | $10^{-23}$ C$^3$J$^{-2}$ | $10^{-33}$ C$^4$mJ$^{-3}$ |
| [N$_{0,0,0,4}$]$^+$ | 0.26 | 3.61 | 4.98 |
| [P$_{4,4,4,4}$]$^+$ | 0.30 | 0.90 | 5.85 |
| [C$_4$mpyrr]$^+$ | 0.28 | 1.26 | 4.00 |
| [C$_4$mpip]$^+$ | 0.30 | 0.66 | 4.19 |
| [C$_4$mmor]$^+$ | 0.31 | 1.42 | 4.22 |
| [C$_4$mim]$^+$ | 0.58 | 2.68 | 5.64 |
| [C$_4$mthia]$^+$ | 0.59 | 4.39 | 7.33 |
| [C$_4$py]$^+$ | 0.59 | 6.96 | 8.22 |
| [C$_4$quin]$^+$ | 0.63 | 1.54 | 5.99 |

## 3. Results and discussion

### 3.1. Influence of the cationic core

A large number of IL-forming cations are derivatives of ammonium and phosphonium cations, or cationic heterocycles bearing alkyl chains of variable length, $k$. In Table 1, the MCBO, M1HD and M2HD of different IL-forming cations with a butyl side chain ($k = 4$) are shown.

6The considered heterocycles can be separated into two groups according to their MCBO values. The first group is formed by aliphatic cations with MCBO values ranging from 0.26 to 0.31, and it includes ammonium and phosphonium derivatives as well as non-conjugated heterocyclic cations. The M1HD and M2HD values in this group are, respectively, within the range of $0.66 - 3.61 \cdot 10^{-23}$ C$^3$J$^{-2}$ and $4.00 - 5.85 \cdot 10^{-33}$ C$^4$mJ$^{-3}$. While [C$_4$mmor]$^+$, [C$_4$mpip]$^+$ and [C$_4$mpyrr]$^+$ heterocyclic cations present similar M2HDs, the M1HD is slightly lower for [C$_4$mpip]$^+$. On the other hand, the ammonium and phosphonium-based cations, [N$_{0,0,0,4}$]$^+$ and [P$_{4,4,4,4}$]$^+$, present the highest M2HDs, behaviour that could be related to their lack of cycles and linear arrangement of their alkyl chains, as discussed later in the text. Since symmetry plays a major role in molecular first hyperpolarizability [54,69], the M1HD of the very asymmetric [N$_{0,0,0,4}$]$^+$ cation is high whereas the M1HD of the centrosymmetric [P$_{4,4,4,4}$]$^+$ cation could be expected to be negligible. Nevertheless, it presents a low but non negligible value, which, according to Ref. [52], could be explained by a certain symmetry distortion originated by the Janh-Teller effect arising from the lack of an electron in the phosphorus atom.

The second group of cations is composed by those that show a certain degree of charge delocalization, their MCBO values range from 0.58 to 0.63, and their M1HD and M2HD are, respectively, within the range of $1.54 - 6.96 \cdot 10^{-23}$ C$^3$J$^{-2}$ and $5.64 - 8.22 \cdot 10^{-33}$ C$^4$mJ$^{-3}$. Both the M1HD and M2HD of the cations bearing a single ring in this group grow in this order: [C$_4$mim]$^+$ < [C$_4$mthia]$^+$ < [C$_4$py]$^+$. With regard to the [C$_4$quin]$^+$ cation, it does not lead to particularly high hyperpolarizability densities, maybe because of the presence of the second cycle in its structure.

### 3.2. Influence of the alkyl chain length

The cations considered up to now are often tailored by modifying the length of their alkyl chains. Fig. 1, shows the evolution of $\beta_{\parallel}$, M1HD, $\langle\gamma\rangle$ and M2HD as a function of the alkyl chain length for the cationic cores considered in the previous section.

The first hyperpolarizability, $\beta_{\parallel}$, and second hyperpolarizability, $\langle\gamma\rangle$, increase with the chain length for all the cations. In both cases increasing the number of CH$_2$ units produces a net contribution to the hyperpolarizabilities, while, for the first hyperpolarizability case, the increase of the asymmetry of the molecule with the chain length may also play a role. With regard to the increase ratio (slopes), it is similar for the aliphatic cations and lower than that of the cations with aromatic heterocycles. For the aromatic cations the first and second hyperpolarizabilities increase in the same order found for the species with $k = 4$, [C$_k$mim]$^+$ < [C$_k$mthia]$^+$ < [C$_k$py]$^+$. Again, an exception to this trend is [C$_k$quin]$^+$, which behaves similarly to an aliphatic cation, maybe, again, due to the second ring. The difference in slopes as a function of the heterocycle conjugation was already observed by Castellanos-Águila et al. [50], when comparing the first hyperpolarizability of pyridinium and imidazolium cations as a function of the chain length. In that work, the slope differences were proposed to be originated by changes in the conjugation of the heterocycles produced by the steric effects associated to the increase of the alkyl chain.

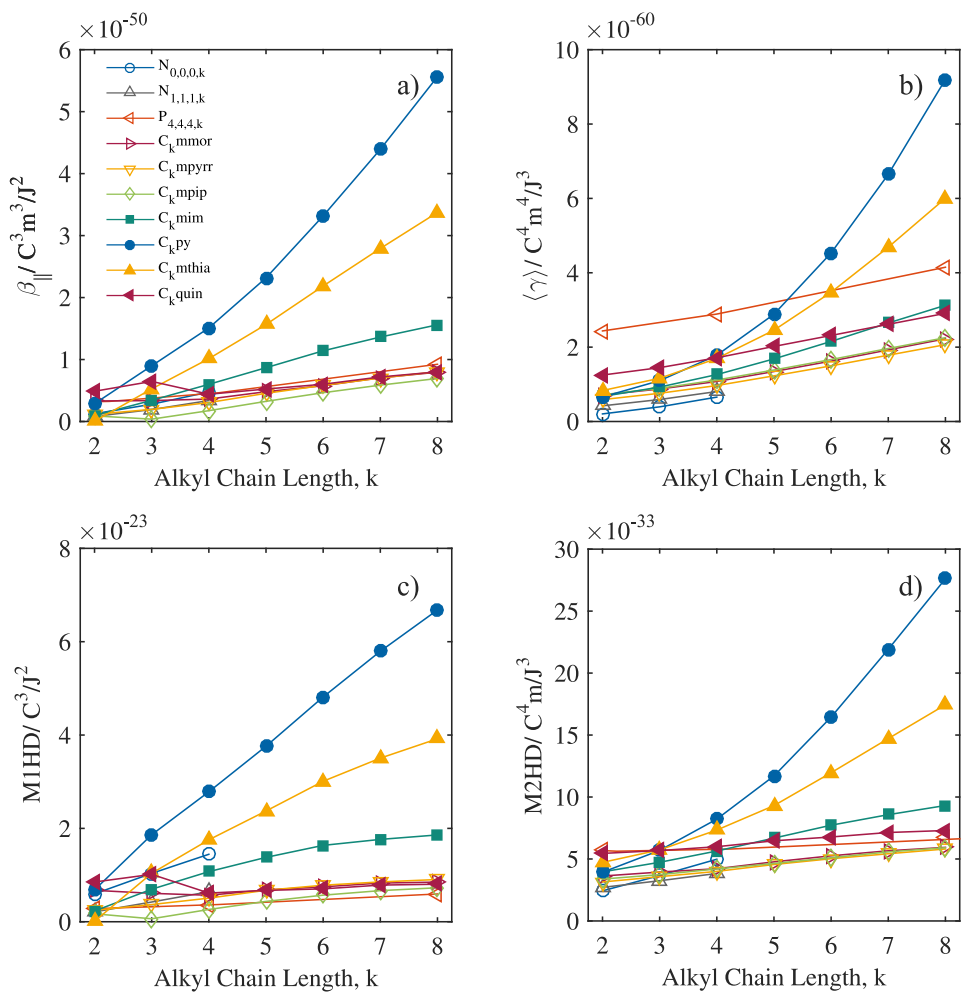

**Fig. 1.** (a) First hyperpolarizability, $\beta_\parallel$, (b) M1HD, (c) second hyperpolarizability $\langle\gamma\rangle$ and (d) M2HD as a function of the alkyl chain length for several ammonium, phosphonium and heterocycle-based cations.

The increase of first and second hyperpolarizabilities with the alkyl chain length is in agreement with the results obtained in recent computational works for $[N_{2,k,k,k}]^+$ based cations [52] as well as imidazolium and pyridinium based cations [49,50,35]. With regard to the numerical agreement of first hyperpolarizability, our figures are slightly higher but compatible with those given in the above mentioned works, specially taken into account that we are comparing with static calculations, sometimes involving smaller basis sets and using different definitions of first hyperpolarizability. For instance, for the imidazolium cation with $k = 4 - 8$, our first hyperpolarizability (see Supplementary Material) is in the range of $1.6 - 4.2 \cdot 10^{-30}$ esu while in Ref. [50] it is approximately in the range of $1.2 - 3.2 \cdot 10^{-30}$ esu. For the pyridinium cation, at the same $k$ interval, we obtained first hyperpolarizability values of $4.1 - 15.0 \cdot 10^{-30}$ esu and, in Ref. [50], it was approximately obtained $2.3 - 4.9 \cdot 10^{-30}$ esu. Moreover, for the $[C_4mim]^+$ cation, our first hyperpolarizability value is slightly higher than that given by Bardak et al. [53] using the M06-2X functional, while, our second hyperpolarizability value, is almost one order of magnitude smaller, showing the strong functional dependence of these magnitudes. Additionally to computational evidences, available experimental measurements on the first hyperpolarizability of imidazolium-based ILs also point to a small increase of this magnitude with the chain length for the $[C_kmim]^+$ family [23].

With respect to the hyperpolarizability densities, values shown by aliphatic cations are smaller than those of the aromatic ones, which increase in the order $[C_kmim]^+ < [C_kmthia]^+ < [C_kpy]^+$. The introduction of the molecular volume compensates the large hyperpolarizabilities of the bulkiest cation families, such as $[P_{4,4,4,k}]^+$, as well as reduces the impact of increasing the alkyl chain length. For long alkyl chains, all aliphatic cations (and $[C_kquin]^+$), reach a saturation regime with similar hyperpolarizability density values, which is not the case for aromatic cations, which benefit in a larger extent of increasing the alkyl chain length.

From these results, it is clear the importance of using aromatic heterocycles to improve the NL optical response of ILs. Specifically, it seems that among the considered cations, the ones derived from the thiazolium or pyridinium heterocycles are the most suitable to be used to design ILs with enhanced NL optical response.

### 3.3. Influence of substituents on the hyperpolarizability

A recurrent mechanism for tuning the properties of IL forming cations is incorporating functionalized side chains. For this reason, we calculated the hyperpolarizability densities of an imidazolium cation bearing one or two substituents with different number of $\pi$ bonds and degrees of charge delocalization. The MCBO, M1HD and M2HD of these cations are shown in Table 2.

Singly substituted imidazoliums present higher hyperpolarizability densities as higher is the MCBO value of the substituent. Doubling the substituent leads to different behaviours for M1HD and M2HD. While the M1HD always decreases, since the symmetry of the cation increases, the M2HD always increases, behaviour in accordance with Eq. (3), which ensures that each new bond contributes to increasing the overall M2HD value. The substituted cations can be distributed in three categories according to their MCBO values.

The first group contains imidazoliums with alkyl, ethoxy and ethylnitrile substituents. The first two lack any $\pi$ bonds, while the third one presents a sp hybridized carbon linked to an N atom. These cations show the lowest hyperpolarizability densities and their MCBO values range from 0.32 to 0.34. Among them, the highest values are those of the butyl group, which also has the largest number of atoms. Comparing propyl, ethoxy and ethylnitrile substituents suggests that substituting C atoms by N or O atoms does not significantly increase the NL response.

The second group comprises carbon-based substituents incorporating $\pi$ bonds in different positions and MCBO values ranging from 0.32 to 0.37. The presence of isolated $\pi$ bonds seems to increase the hyperpolarizability density with respect to $\sigma$ bonds. It can be observed in the higher value of M2HD for [allylmim]$^+$ and [crotylmim]$^+$ cations with respect to their saturated equivalents, [C$_3$mim]$^+$ and [C$_4$mim]$^+$. Curiously, this trend is not totally true for M1HD since the [allylmim]$^+$ cation shows a slightly lower value than [C$_3$mim]$^+$. The benzyl group presents an extensive $\pi$ conjugated system and, for this reason, it is expected to present large hyperpolarizability densities. Indeed, according to Eq. (3), the effective $L$ for the benzene ring is larger than that of the rest of substituents within this group.

The third group comprises imidazoliums with different number of vinyl groups attached to different positions. This category is the one showing the highest MCBO values, ranging from 0.39 to 0.48. The high values of the MCBO indices suggest that the electrons in the vinyl group participate to a certain extent in the charge delocalization throughout the imidazolium ring [78]. The M1HDs in this category are not very large due to the large degree of symmetry of the considered cations. As expected, introducing a second vinyl group in the second N of the imidazolium ring produces an important decrease of the overall M1HD. However, interestingly, introducing it in the C between both N atoms ([1,2-divinylmim]$^+$) yields to a much limited reduction of M1HD, and specially, leads to change in the sign of the first hyperpolarizability value, which becomes negative. The negative sign in [1,2-divinylmim]$^+$ suggests a net electron flow from the imidazolium ring to the vinyl group, which is playing an electron acceptor role [79]. On the other hand, the M2HD of the vinyl-substituted imidazolium is large and doubling it produces the largest M2HD of the studied substituents. Furthermore, the linkage of the vinyl group to the N or the C atoms has different effect on the overall M2HD, which is another evidence of the affectation of electronic conjugation throughout the imidazolium ring on hyperpolarizability.

A noteworthy important fact is that the increase of hyperpolarizability densities due to rings is not as high as expected when comparing it with that of linear chains. Focusing on M2HDs to leave aside strong symmetry effects, aromatic cyclic substituents, such as the benzyl group, which contains a cycle of 6 conjugated carbon atoms, present NL responses not much higher than those of much shorter linear conjugated substituents, such as those of the vinyl group. Similar evidences of the influence of linear and cyclic arragements of atoms in the hyperpolarizabilities of cations were also observed for the ammonium, phosphonium and quinolinium charged moieties of the previous section. The differences between the intensity in the NL response of linear and cyclic charge distributions are related to the existence of a dimensional effect [80], which is widely supported by available experimental measurements [55,81].

### 3.4. Influence of functionalized chains on the hyperpolarizability

Atomic arrangement and charge delocalization by means of delocalized $\pi$ bonds play a decisive role in the NL response of cations. For this reason, four families of imidazoliums bearing functionalized side chains of variable length were simulated. The side chains considered are i) a perfluorinated alkyl chain where the H atoms are replaced by F atoms, ii) an oxygenated chain where the repetition unit is an oxyethyl group, iii) a polyenylic chain where there is an alternance of single and double bonds between carbons, and iv) a common alkyl chain. The first two are interesting since they content chemical elements beyond H and C. The last two present totally opposite characters since the alkyl chain only contains highly localized $\sigma$ bonds while the polynenyl chain contains resonant $\pi$ bonds that delocalize the charge throughout all the chain length. Fig. 2 shows the M1HD and M2HD of the different types of side chains as a function of their length.

Increasing the chain length leads to a increase of the M1HD values, which present a minimum at the chain length providing cations with the highest symmetry. Alkyl, perfluorinated and oxyethyl chains yield moderate increases of M1HD with the chain length, with the last one showing a saturation effect for chain lengths larger than $k = 2$. Among them, the alkyl chain is the one which potentially provides the larger NL response by increasing the chain length. However, these responses are much weaker than that of the polynelylic chain, which produces a huge increase of the M1HD of the cation.

The cations bearing alkyl, perfluorinated and oxyethyl chains show moderate increases of M2HD with $k$, but much smaller than that of the polyenylic chain. The comparison of the alkyl chain with those containing heteroatoms suggests that neither the introduction of F or O in the molecular moieties is positive for achieving larger NL responses. Interestingly, the introduction of the first unit of the perfluorinated chain is detrimental for the NL response of the imidazolium ring since it decreases from $k = 0$ to $k = 1$, afterwards, it continuously increases with $k$. Moreover, the M2HD of the oxygenated chain seems to reach saturation for relatively low $k$ values. Indeed, for $k = 5$ the alkyl chain length reaches the NL response of the oxygenated chain, even if this last one contains two carbon atoms and one oxygen atom per $k$ unit.

### 3.5. Influence of charge delocalization regions of variable length on the hyperpolarizability

In order to obtain a closer insight on the origins of the differences found for both M1HDs and M2HDs in chains containing only $\sigma$ bonds and extended $\pi$ conjugated systems, the molecular volume, $V$, first hyperpolarizability, $\beta_{\parallel}$, and second hyperpolarizability, $\langle\gamma\rangle$, of an imidazolium cation bearing a saturated alkyl chain or an unsaturated (conjugated) polyenyl chain are shown in Fig. 3.

**Table 2**
Normalized MCBO indices, M1HD and M2HD of imidazoliums having different substituents.

| Cation | \|MCBO\| | M1HD $10^{-23}$ C$^3$J$^{-2}$ | M2HD $10^{-33}$ C$^4$mJ$^{-3}$ |
|---|---|---|---|
| [C$_3$mim]$^+$ | 0.33 | 1.74 | 4.73 |
| [C$_4$mim]$^+$ | 0.34 | 2.68 | 5.64 |
| [eomim]$^+$ | 0.33 | 1.83 | 4.14 |
| [1,3-dieoim]$^+$ | 0.34 | 0.78 | 4.67 |
| [enmim]$^+$ | 0.34 | 0.36 | 3.70 |
| [1,3-dienim]$^+$ | 0.32 | 0.26 | 3.90 |
| [allylmim]$^+$ | 0.35 | 1.12 | 5.34 |
| [1,3-diallylim]$^+$ | 0.37 | 0.14 | 6.33 |
| [crotylmim]$^+$ | 0.32 | 3.60 | 6.42 |
| [benzylmim]$^+$ | 0.36 | 3.28 | 7.32 |
| [1,3-dibenzylim]$^+$ | 0.35 | 1.19 | 9.67 |
| [vinylmim]$^+$ | 0.40 | 1.26 | 5.84 |
| [1,2-divinylmim]$^+$ | 0.39 | -1.17 | 7.55 |
| [1,3-divinylim]$^+$ | 0.48 | 0.67 | 11.39 |

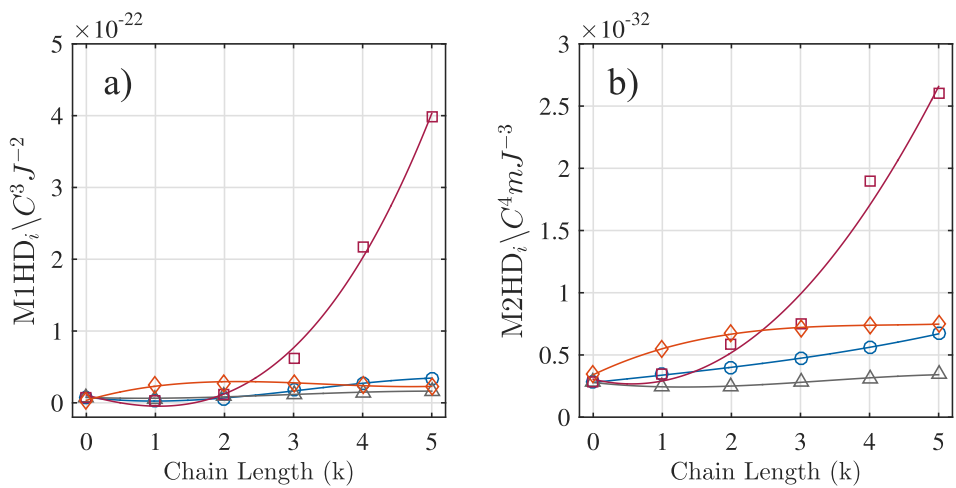

**Fig. 2.** M1HDs and M2HDs of imidazolium cations with side chains of variable length. Marker legend: -CH$_2$- (○), -CF$_2$- (△), -CH= (or = CH-) (□) and -CH$_2$-O-CH$_2$- (◇). Solid lines only mean to be a guide to the eye.

Molecular volume is linear upon increase of both chains but the slope is larger in the saturated case than for the unsaturated case, which is explained by the shorter bond distance and the lack of a hydrogen atom in its basic unit. This volume inspection reveals that even if the basic units of both chains had the same contribution to hyperpolarizability, the unsaturated one would present larger M1HDs and M2HDs values due to its smaller volume. Nevertheless, conjugated chains also present larger contributions to hyperpolarizabilities than saturated alkyl chains, whose increase is roughly linear with $k$. Leaving aside geometrical effects, the different behaviours can be qualitatively explained by means of the simple model of Eq. (3). In a saturated chain, each $\sigma$ bond is an independent region of highly localized charge of length $L$ with an independent contribution to the overall molecular hyperpolarizability. This explains the approximately linear increase of $\beta_{\parallel}$ and $\langle\gamma\rangle$ with $k$, which also was experimentally observed for other families of molecules [82,60,83]. Oppositely, in the polyenylic chain, each new $\pi$ bond contributes to expand the $\pi$ conjugated system, thus, there is an effective increase of the length $L$ in Eq. (3), where the delocalized electrons are allowed to move. This effect yields a NL increase of the hyperpolarizability due to the power dependence on $L$. The resulting difference of contributions of alkyl and polyenyl chains to hyperpolarizabilities is huge. For instance, for $k=5$, the values of $\beta_{\parallel}$ and $\langle\gamma\rangle$ of the cation bearing the polyenyl chain are respectively about 10.5 and 3.5 times larger than those of the saturated chain. Hence, the introduction of linear conjugated regions is a promising mechanism to enhance the NL optical response of specific IL-forming cations.

### 3.6. Charge delocalization and hyperpolarizability of anions

Anions commonly used in ILs generally do not present extensive conjugated systems. However, they have $\pi$ orbitals since they often present double and triple bonds. Some examples of families of anions having double bonds are acetates, triflates or sulfates while triple bonds are present in all anions with the nitrile group such as dicyanamide or thiocyanate. The $\pi$ orbitals participating in these double and triple bonds often form part of small resonant structures that cover an important part of the anions. Hence, despite most of the commonly used anions do not have large conjugated systems, some of them present charge delocalization on considerably large fractions of their molecular structures. The MCBO index, M1HD and M2HD of the studied IL-forming anions are shown in Table 3.

MCBO indices of the anions range from 0.17 to 0.73 due to their large structural and compositional differences. Anions with MCBO ranging from 0.17 to 0.31 show, in general, the lowest M1HD and M2HDs. The M1HDs are very low in magnitude and of negative sign for [OTf]$^-$ and [NTf$_2$]$^-$, and even negligible, for the more symmetric [B(CN)$_4$]$^-$ anion. The highest M1HD within this group is found for [C$_1$SO$_4$]$^-$, which also shows the largest MCBO. Anions with MCBO indices in the range from 0.40 to 0.49 show higher hyperpolarizability densities than those of lower MCBO indices. The exception is $[N(CN)_2]^-$ which shows slightly lower values for both M1HD and M2HD than $[C_1SO_4]^-$ despite having a larger MCBO index. In this case, the M1HD is negative but of greater magnitude than those shown by $[OTf]^-$ or $[NTf_2]^-$. For the other anions in this MCBO range, the increase of hyperpolarizability density is evident. The anions with MCBO values comprised between 0.51 and 0.73 show the highest hyperpolarizabilities densities, being the highest ones those of the $[Oac]^-$ anion. The M1HDs and M2HDs of all the studied anions are shown in Fig. 4.

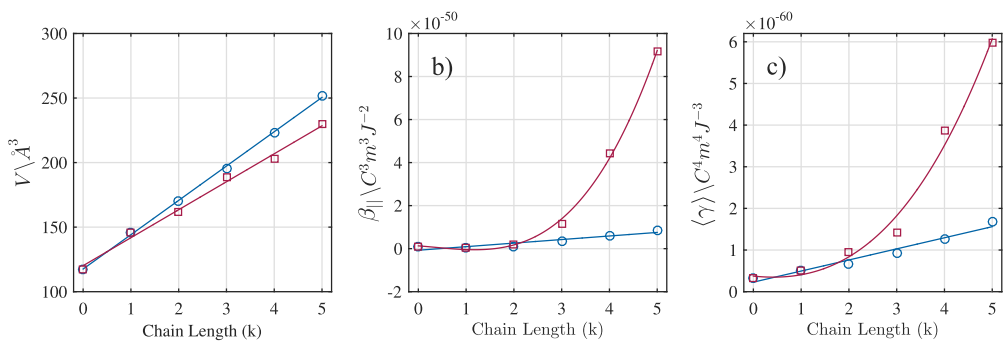

**Fig. 3.** a) Molecular volume, V, b) first hyperpolarizability for SHG, $\beta_{\parallel}$, and, c) second hyperpolarizability for EOKE, $\langle\gamma\rangle$, of a saturated alkyl chain (○) and a polynenyl (conjugated) chain (□) of variable length, $k$. Solid lines are guides to the eye.

As expected, symmetric anions such as $[BF_4]^-$, $[PF_6]^-$, $[AlCl_4]^-$, $[FAP]^-$ or the above mentioned $[B(CN)_4]^-$ present a vanishing first hyperpolarizability density, which is also true for mono-atomic anions such as $[Cl]^-$ or $[Br]^-$. Highly fluorinated anions tend to provide small but negative first hyperpolarizabilities as it is the case of $[Oaf]^-$, $[OTf]^-$ or $[NTf_2]^-$ while asymmetric anions lacking fluorine provide positive first hyperpolarizability and higher M1HDs as it is the case of $[C_1SO_3]^-$, $[C_1SO_4]^-$, $[Tos]^-$, $[SCN]^-$ or $[Oac]^-$. It is interesting how the fluorination of the $[Oac]^-$ to yield $[Oaf]^-$, apart from changing the hyperpolarizability sign, produces a strong reduction of its magnitude, already observed by Bardak et al. [53]. Decrease of the M1HD also happens upon introduction of a new O atom in $[C_1SO_3]^-$ to produce $[C_1SO_4]^-$, indicating that introduction of high electronegative atoms could be counter-productive if not used properly. In this regard, it is well known that combining donor and acceptor regions in a molecule linked by extended conjugated systems greatly enhances hyperpolarizabilities [54].

The trend for the first hyperpolarizabilities used to calculate the M1HD values (see Supplementary Material) matches that of isolated anions in other computational works [49,53], which also predict negligible first hyperpolarizability for symmetric anions and good responses for anions showing charge delocalization such as $[Oaf]^-$, $[Tos]^-$ or $[Oac]^-$. At the quantitative level agreement is not so good, since very different levels of theory were used in the calculations. Our values are up to three times higher than those at the M06-2X/6–311++G(d,p) level of theory shown in Ref. [53]. Differences in signs could be attributed to the usage of different conventions and different functionals.

Regarding M2HDs, the smallest values are shown by anions containing a large amount of fluorine atoms such as $[BF_4]^-$, $[PF_6]^-$ or also $[FAP]^-$, which again shows that the introduction of fluorine is detrimental. Other organic anions containing fluorine such as $[OTf]^-$ or $[NTf_2]^-$ present slightly higher M2HD values. Mono-atomic halogen anions show relatively good M2HDs, which seem to be larger as larger is the atomic number of the atom. Introduction of metal atoms by means of metallic complexes such as $[AlCl_4]^-$, produces moderate responses, at least as far as closed-shell metallic cations are involved. However, the largest responses are those of organic anions lacking of fluorine, such as $[C_1SO_3]^-$, $[C_1SO_4]^-$, $[Tos]^-$, $[SCN]^-$ or $[Oac]^-$. Again, fluorination of $[Oac]^-$ to yield $[Oaf]^-$ decreases the M2HDs as well as introduction of a fourth oxygen atom in $[C_1SO_3]^-$ to produce $[C_1SO_4]^-$. Curiously, despite the extensive conjugated system in $[Tos]^-$, its M2HD is not the highest one, possibly due to cyclic arrange of this moiety. The large values of M1HDs and M2HDs of $[SCN]^-$ and $[Oac]^-$ make them the most promising anions to achieve strong NL responses in ILs.

**Table 3**
MCBO, M1HD and M2HD of selected anions.

| Anion | \|MCBO\| | M1HD $10^{-23}$ $C^3J^{-2}$ | M2HD $10^{-33}$ $C^4mJ^{-3}$ |
|---|---|---|---|
| $[OTf]^-$ | 0.17 | −0.56 | 6.01 |
| $[NTf_2]^-$ | 0.23 | −0.34 | 5.02 |
| $[B(CN)_4]^-$ | 0.26 | $\simeq -0$ | 5.65 |
| $[C_1SO_4]^-$ | 0.31 | 5.07 | 9.55 |
| $[Tos]^-$ | 0.40 | 9.27 | 15.87 |
| $[C_1SO_3]^-$ | 0.47 | 11.44 | 17.31 |
| $[N(CN)_2]^-$ | 0.49 | −3.48 | 9.26 |
| $[Oac]^-$ | 0.51 | 53.59 | 72.72 |
| $[SCN]^-$ | 0.73 | 13.41 | 27.14 |

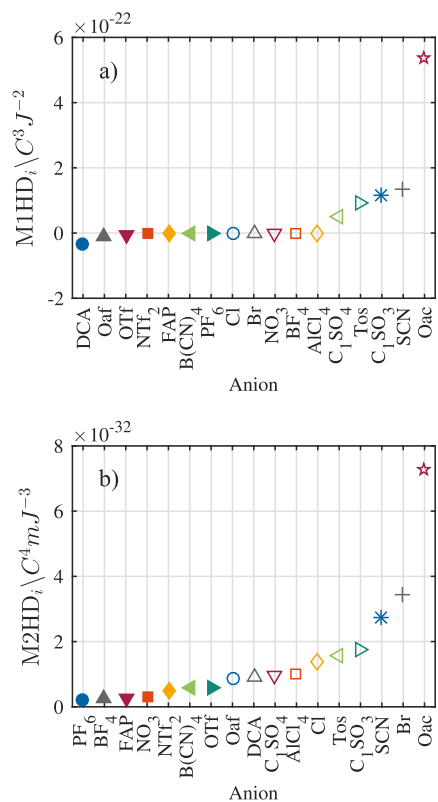

**Fig. 4.** a) M1HD and b) M2HD for the studied anions.

The second hyperpolarizability used to calculate the M2HD values (see Supplementary Material) shows a trend almost identical to that provided in Ref. [53] for isolated anions at the M06-2X/6–311++G(d,p) level of theory. However, the numerical agreement is not good because of the differences in the chosen functionals. Our values are one order of magnitude smaller than those presented in that work, just the opposite to the behaviour found for the first hyperpolarizabilities. Interestingly, the trend we predict for anions is also close to that shown in Ref. [53] for ILs sharing different anions with the same cation, which points to the large influence of anions in the second hyperpolarizabilities of ILs.

### 3.7. Additivity of Hyperpolarizabilities

The non-linear susceptibilities of ILs can be estimated by means of Eqs. 7 and 8. However, in order to do that, the hyperpolarizability of the ionic pair must be known. The most common procedure to calculate it is performing joint simulations of ionic pairs [49,50,52,53], or even considering a higher number of ionic dimers, [23]. An important conclusion of these works is that charge transfer between ions plays a relevant role in the overall hyperpolarizability of ILs. We have simulated the first and second hyperpolarizability of 36 different ionic pairs together and compared the results with those obtained from the linear addition of the contributions of the $i$ ions they are made of, $\beta_{IL} = \sum_i^N \beta_{\|,i}$ and $\gamma_{IL} = \sum_i^N \langle \gamma_i \rangle$. The considered pairs were combinations of ions with different extents of conjugation and symmetries: three cationic heterocycles, $[C_k\text{mpyrr}]^+$, $[C_k\text{mim}]^+$ and $[C_k\text{py}]^+$, with three differ-ent alkyl chain lengths, $k = 2, 4, 8$, and four anions, [SCN]$^-$, [Oaf]$^-$, [NTf$_2$]$^-$ and [BF$_4$]$^-$. The results are shown in Fig. 5.

First hyperpolarizabilities are largely non-additive, in perfect agreement with [49,50,52,53], but second hyperpolarizabilities present a region where additivity holds, behaviour that was already noted in Ref. [53]. The strong differences in the linearity of both hyperpolarizabilities could be related to the important dependence on geometry shown by the first hyperpolarizability, largely affected by the differences in the symmetry of isolated ions and ionic pairs. With respect to second hyperpolarizability, linearity seems to hold for most of the tested ion combinations, but it partially breaks down when dealing with ILs containing highly resonant anions such as [SCN]$^-$, and in minor extent, [Oaf]$^-$. This fact could indicate that non-linearities are stronger in the presence of highly delocalized charges but also a limitation of the chosen functional to properly describe these interactions. Further research is required to shed more light on this point. Nevertheless, even in the cases where the linearity weakens, the differences between the second hyperpolarizability calculated by both methods are, in general, below than a factor of 2.

### 3.8. Non linear susceptibilities of ILs

Using Eqs. 7 and 8, the non linear susceptibility of any IL can be calculated. The field factor can be estimated using the refractive index of the IL at $\lambda$= 1100 nm. The required refractive index and molecular volume of the IL can be easily calculated from that of the isolated ions following the procedure described in Ref. [78]. This procedure assumes additivity in the polarizability and that the IL molecular volume is obtained as $V_{IL} = f_{scale} \cdot \sum_i V_i + f_{int} \cdot (\sum_i V_i)^2 > \sum_i V_i$, being $f_{scale} = 1.0189$ and $f_{int} = 3.635 \times 10^{-4}$ Å$^{-3}$ semi-empirical parameters optimized for volume calculations at the B3LYP/6–31++G(d,p) level of theory. On the other hand, the hyperpolarizability of the IL, at least the second order one, can be approximately calculated as the sum of contributions of its ions. This assumption provides a computationally affordable and time-saving strategy to identify the most interesting susceptibilities for NL optics in the IL world. Since additivity is not ensured for first hyperpolarizability, the rest of this section is focused on third order susceptibility, but a similar discussion for second order susceptibility was included in the Supplementary Material.

Under the linearity assumption of second order hyperpolarizability, the third order susceptibility of an IL, $\chi^{(3)}$, can be rationalized in terms of the M2HDs of its composing ions:

$$\chi^{(3)} \propto \frac{\gamma_{IL}}{V_{IL}} \leqslant \sum_i^N \text{M2HD}_i \cdot \phi_i, \quad (9)$$

being $\phi_i = V_i/\sum_i V_i$ the volume fraction of each species. The inequality mean that the IL susceptibility would be lower than that of the ideal mixture, since the effective volume of the IL is larger than the sum of the ions. According to Eq. 9, it could seem that the third order NL susceptibilities of ILs are governed by the commonly bulky cations. Nevertheless, it is important to note that small anions can present much higher hyperpolarizability densities, counterbalancing the cationic dominance in terms of volume, Table 4.

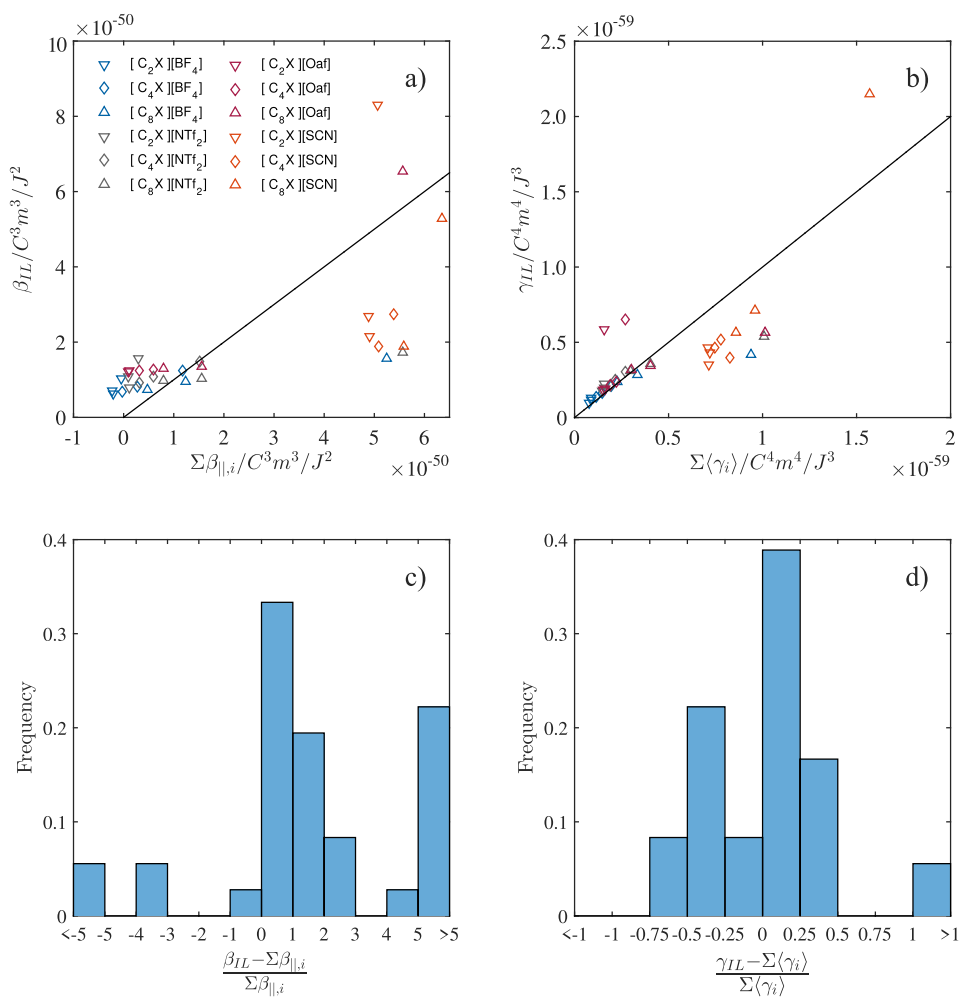

**Fig. 5.** Comparison between first hyperpolarizability (a) and second hyperpolarizability (b) of 36 ILs calculated as the sum of the hyperpolarizabilities of the ions, $\sum_i \beta_{\|,i}$ and $\sum_i \langle \gamma_i \rangle$, and those given by DFT calculations over the ionic pairs, $\beta_{IL}$ and $\gamma_{IL}$. Relative frequency histograms show the relative deviation of both methods in the calculation of (c) first hyperpolarizability and (d) second hyperpolarizability as a function of the number of ILs showing that relative deviation.

Indeed, the third order NL susceptibility of an IL is given by a delicate interplay between the contributions of its constituent ions. For instance, commonly used inorganic anions such as $[PF_6]^-$ or $[BF_4]^-$ present lower values than widely used cations such as $[C_4mim]^+$, $[C_4mpip]^+$ or $[C_4py]^+$. However, other frequent anions such as $[OTf]^-$ or $[NTf_2]^-$ present responses similar to those of these cations, and other anions such as $[Tos]^-$, $[SCN]^-$ or $[Oac]^-$, show much higher M2HDs. In general, a large number of ion combinations yield M2HDs higher than that of the urea, molecule often taken as reference in the NL optics field, whose M2HD is $4.07 \cdot 10^{-33}$ $C^4mJ^{-3}$ [84]. The $\chi^3(-\omega;\omega,0,0)$ susceptibilities for the EOKE process at $\lambda$=1100 nm, for all the combinations of the ions considered in this work are shown in the color map of Fig. 6. The corresponding values can be consulted in the Supplementary Material.

The figure shows that, combinations including the most common IL-forming anions such as $[BF_4]^-$ $[OTf]^-$ or $[NTf_2]^-$, are detrimental to obtain high third order susceptibilities. Nevertheless, combinations involving anions and cations with large degrees of charge delocalization, i.e., those including aromatic or resonant anions such as $[Tos]^-$, $[SCN]^-$ or $[Oac]^-$ together with strongly aromatic heterocyclic cations such as $[C_4thia]^+$, $[C_4py]^+$ produce the highest third order susceptibilities. Further tailoring of the cation, such as the introduction of a conjugated chain, as in the $[uC_kmim]^+$ family, produces a great increase of the final third susceptibility of the IL, reaching values for $k = 5$ as high as $277 \cdot 10^{-22}$ $m^2/V^2$. However, further research on these combinations of highly delocalized ions are required, since our additive model is less accurate in this region of the susceptibility map. In this regard, some works suggest that long range charge transfer [85] and thus, accurate predictions of hyperpolarizabilities [86], only can be properly achieved by the employment of the long-range corrected functionals [87] such as CAM-B3LYP [88] or LC-$\omega$PBE [89].

Comparison of our results with those of other theoretical works can be done in terms of the second hyperpolarizability since susceptibilities are not frequently simulated. For ILs based on combinations of the $[C_4mim]^+$ cation with different anions, we predict a trend of second hyperpolarizability, which is in qualitative agreement with that of the calculations of Ref. [53] at the M06-2X/6–311++G(d,p) level of theory. The most important discrepancies are found for the $[C_4mim][Br]$ and $[C_4mim][Oac]$ ILs, since we predict for this last IL the highest second hyperpolarizability of the series while this place is occupied by $[C_4mim][Br]$ in Ref. [53]. At the quantitative level, our values are closely one order magnitude smaller than those presented in that work. Nevertheless, this is not surprising since this offset was already noted for isolated ions. Higher quantitative agreement is found for the second hyperpolarizability of the $[C_4mim][Oaf]$ given in Ref. [35], which is just a 20%

**Table 4**
M2HD of selected anions and cations.

| Anion | M2HD $10^{-33}$ $C^4mJ^{-3}$ | Cation | M2HD $10^{-33}$ $C^4mJ^{-3}$ |
|---|---|---|---|
| $[PF_6]^-$ | 1.91 | $[F_5mim]^+$ | 3.42 |
| $[BF_4]^-$ | 2.58 | $[C_4mpip]^+$ | 4.19 |
| $[NTf_2]^-$ | 5.02 | $[C_4mim]^+$ | 5.64 |
| $[OTf]^-$ | 6.01 | $[vinylmim]^+$ | 5.84 |
| $[N(CN)_2]^-$ | 9.26 | $[m(eo)_5mim]^+$ | 7.47 |
| $[Tos]^-$ | 15.87 | $[C_4py]^+$ | 8.22 |
| $[SCN]^-$ | 27.14 | $[1,3$-divinylim$]^+$ | 11.39 |
| $[Oac]^-$ | 72.72 | $[uC_5mim]^+$ | 26.47 |

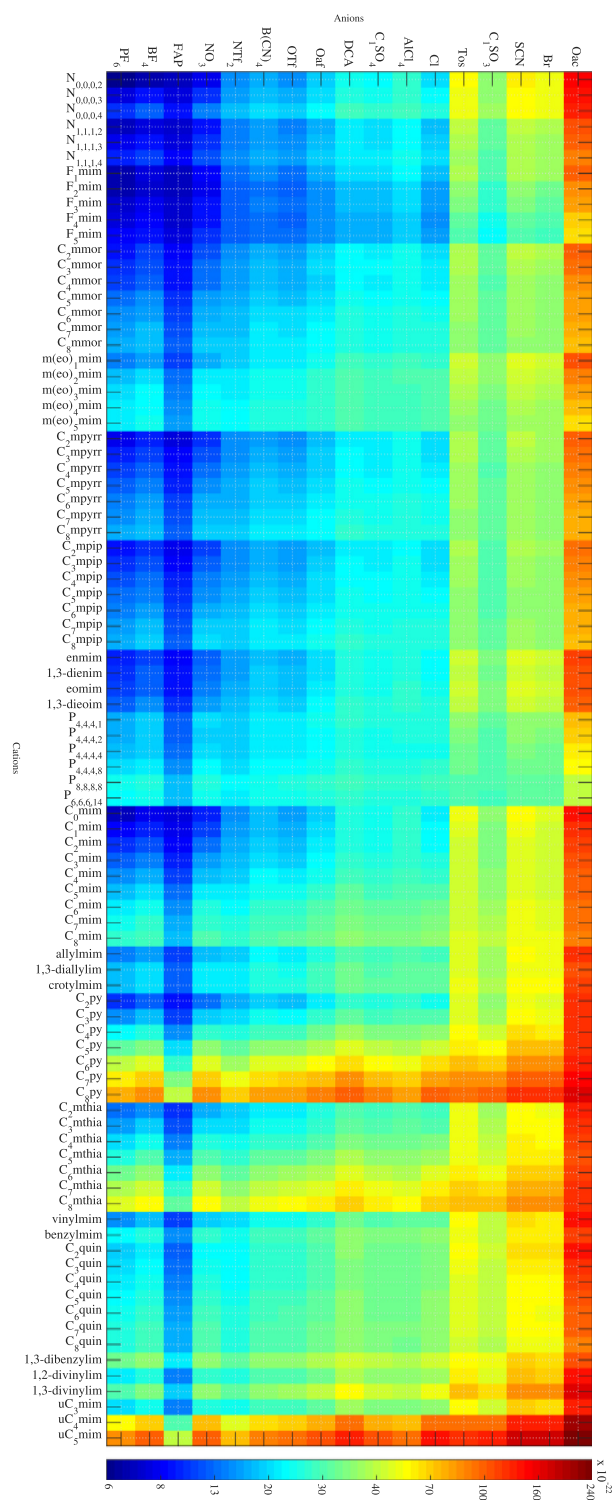

**Fig. 6.** Color map of the third order susceptibility $\chi^3(-\omega;\omega,0,0)$ for EOKE at $\lambda$=1100 nm for the combinations of all the studied ions. SI units, $1 \cdot 10^{-22}$ m$^2$ V$^{-2}$ and color in logarithmic scale.

larger than the one in this work. In relation to experimental measurements, available data indicate that our simulations of second order susceptibility overestimate that of the experiments in one order of magnitude. For instance, using the simulated refractive indices of [78] and the experimental measurements of [34], we estimate an experimental third order susceptibility for $[m(eo)_1$-mim$][NTf_2]$ of $0.15 \cdot 10^{-21}$ m$^2/V^2$, while the value simulated in this work is $1.7 \cdot 10^{-21}$ m$^2/V^2$. This trend can also be seen in $[C_4mim]$

C.D. Rodríguez-Fernández, L.M. Varela, C. Schöder et al.  *Journal of Molecular Liquids 349 (2022) 118153*

[OTf], where the experimental third order susceptibility [35] is $2.3 \cdot 10^{-22}$ m$^2$/V$^2$ whereas our estimation yields $22.12 \cdot 10^{-22}$ m$^2$/V$^2$. In this last work, from the experimental third order susceptibility, the experimental second hyperpolarizability was extracted, which also was one order of magnitude smaller than our estimations. Indeed, our calculations were in agreement with those carried out in that work, which also overestimated the experimental second hyperpolarizability in one order of magnitude. The generalized overestimation of the second hyperpolarizability of the B3LYP functional suggests that further investigation of the influence of DFT functionals on the hyperpolarizability of ILs is required.

## 4. Conclusions

ILs are highly versatile materials that can be tailored to produce important non-linear optical responses. The two key factors for designing ILs with enhanced non-linear responses are charge delocalization and molecular symmetry. While charge delocalization is important both for second and third order responses, the symmetry is more relevant for second order responses, since the more symmetric is a molecule, the less elements in its first hyperpolarizability tensor are different from zero. First hyperpolarizabilities of ILs cannot be obtained by simple addition of those of their ions. However, the second hyperpolarizability presents a certain degree of linearity -specifically for ions with low degrees of charge delocalization- which can be exploited to perform efficient calculations of the hyperpolarizability of a large number of ILs. In this regard, we present in this work a map of third order non linear susceptibilities in the IL world which is expected to be a valuable tool to guide future works on the non linear optical properties of ILs. This map indicates that the best option to enhance the non-linear susceptibility of ILs is keeping small resonant anions and tailoring high aromatic cations, attaching them, for instance, substituents including aromatic cycles or long linear conjugated regions.

**Declaration of Competing Interest**

The authors declare that they have no known competing financial interests or personal relationships that could have appeared to influence the work reported in this paper.

**Acknowledgements**

This work was supported by Ministerio de Economia y Competitividad (MINECO) and FEDER 17 Program through the projects (MAT2017-89239-C2-1-P); Xunta de Galicia and FEDER (GRC ED431C 2016/001, ED431D 2017/06, ED431E2018/08). C. D. R. F. thanks the support of Xunta de Galicia through the grant ED481A-2018/032. We also thank the Centro de Supercomputacion de Galicia (CESGA) facility, Santiago de Compostela, Galicia, Spain, for providing the computational resources employed in this work.

**Appendix A. Supplementary material**

Supplementary data associated with this article can be found, in the online version, at https://doi.org/10.1016/j.molliq.2021.118153.